\documentclass[11pt]{article}

\usepackage{graphicx}
\usepackage{amsmath,amssymb,array}
    \RequirePackage{fancyvrb}
    \RequirePackage{alltt}
    \DefineVerbatimEnvironment{example}{Verbatim}{}
    \renewenvironment{example*}{\begin{alltt}}{\end{alltt}}
\usepackage[utf8]{inputenc}
\usepackage{color}

\usepackage[top=2cm, bottom=2cm, left=2cm, right=2cm]{geometry}

\title{\textbf{multiColl}: An R package to detect multicollinearity}

\author{Román Salmerón \\
              Department of Quantitative Methods for the Economy and Business \\
              University of Granada (Spain) \\
              Tel.: +34-958248791\\
              email: romansg@ugr.es \\
           Catalina García  \\
              Department of Quantitative Methods for the Economy and Business \\
              University of Granada (Spain) \\
              Tel.: +34-958248790\\
              email: cbgarcia@ugr.es \\
           José García \\
              Department of Economics and Business \\
              University of Almería (Spain) \\
              Tel.: +34-958248790\\
              email: jgarcia@ual.es
}

\begin{document}

\maketitle

\begin{abstract}
    This work presents a guide for the use of some of the functions of the \textbf{R} \cite{R} package \textbf{multiColl} for the detection of near multicollinearity.  The main contribution, in comparison to other existing packages in \textbf{R} \cite{R} or other econometric software, is the treatment of qualitative independent variables and the intercept in the simple/multiple linear regression model.
\end{abstract}


\section{Introduction}

As is known, linear regression is a widely applied statistical tool for analyzing the effect (if any) of a set of independent variables on a dependent variable, allowing the numerical estimation of signs and magnitudes of coefficients in a previously established linear regression. The multiple regression model is one of the most frequently applied models for establishing a linear relation among variables. A model with $n$ observations and $k$ independent variables is specified as follows:
\begin{equation}
\label{model0}
\mathbf{y}_{n \times 1} = \mathbf{X}_{n \times k} \cdot \boldsymbol{\beta}_{k \times 1} + \mathbf{u}_{n \times 1},
\end{equation}
where the first column of $\mathbf{X}$ is composed of ones representing the intercept (denoted by $\mathbf{1}_{n \times 1} = (1,\dots,1)^{t}$) and $\mathbf{u}$ represents the random disturbance assumed to be centered and spherical\footnote{That is, $E[\mathbf{u}_{n \times 1}] = \mathbf{0}_{n \times 1}$ and $var(\mathbf{u}_{n \times 1}) = \sigma^{2} \cdot \mathbf{I}_{n \times n}$, where $\mathbf{0}$ is a vector of zeros, $\sigma^{2}$ is the variance of the random disturbance, and $\mathbf{I}$ is the identity matrix.}. The aim is to estimate the coefficients $\boldsymbol{\beta}$ of the independent variables and, from these values, establish the direction of relations (according to the signs) and quantify relations (with values).

The ordinary least squares (OLS) approach is the most frequently applied methodology for obtaining the estimates of coefficients using the well-known expression\footnote{It is known that this expression matches the expression given by the method of maximum likelihood.} $\widehat{\boldsymbol{\beta}} = \left( \mathbf{X}^{t} \mathbf{X} \right)^{-1} \mathbf{X}^{t} \mathbf{y}$. Thus, the inverse of matrix $\mathbf{X}^{t} \mathbf{X}$ is required, and, consequently, it is assumed that the variables in matrix $\mathbf{X}$ have a linear relationship and are independent.

If this condition does not hold, it is said that the model is characterized by multicollinearity. Thus, multicollinearity describes the absence of orthogonality between the independent variables of the model.

If multicollinearity is exact, which occurs if one of the independent variables is a perfect linear combination of some or all of the other variables, it is not possible to obtain the inverse of matrix $\mathbf{X}^{t} \mathbf{X}$ and, in this case, the aim will be unattainable, since there will not be a unique estimate of $\widehat{\boldsymbol{\beta}}$. However, if multicollinearity is near, which occurs if one of the independent variables is approximately equal to a linear combination of some or all of the other variables, it is possible to obtain the inverse of the matrix, and consequently a unique estimate of coefficients.

In the first case (perfect multicollinearity), the model does not satisfy the condition of full rank; as previously noted, this fact leads to infinite estimates of coefficients of the regression model. In the second case (near multicollinearity), although the full rank condition is satisfied, the estimates will be unstable, and it is possible to note the following problems related to model estimation and its statistical analysis:
\begin{itemize}
\item Small changes in the data involve large changes in the estimates of coefficients of independent variables in the model; as a consequence, unexpected signs of relations can be obtained.
\item Estimated variances of the estimated coefficients are inflated, so there will be a tendency not to reject the null hypothesis in the analyses of individual significance.
\item There is a tendency to reject the null hypothesis in analyses of joint significance, which will contradict the previously noted results of individual inference.
\end{itemize}

Then, the second case is the problematic one, since the conclusions obtained from the analysis will be questioned; such concerns justify development of an appropriate method to detect and treat collinearity.

On the other hand, the causes of approximate multicollinearity in a model are diverse; however, they can be divided Marquardt and Snee \cite{MarquardtSnee:1975} into the following types:
\begin{description}
\item[Nonessential multicollinearity:] exists due to the relation between the intercept and the rest of the independent variables.
\item[Essential multicollinearity:] exists due to the relation between the independent variables apart from the intercept.
\end{description}

This distinction is useful when applying the most common measures of approximate multicollinearity detection presented below, since it be shown that not all measures are useful for detecting both types of approximate multicollinearity.

Finally, note that some measures are appropriate or not depending on the type of independent variables. For example, it only makes sense to calculate the matrix of simple correlations or Variance Inflation Factors for independent quantitative variables; hence, if, for example, there are dummy variables, they should be ignored in the calculation of these measures.

It is important to emphasize this last aspect, since, to the best of our knowledge, the packages existing up to now in \textbf{R}, as \textbf{car} \cite{car2018}, \textbf{mctest} \cite{mctest2018} or \textbf{perturb} \cite{perturb2012}, do not make any distinction between the nature of independent variables. That is, in these cases, incorrect detection of the approximate multicollinearity would occur.

\section{Multicollinearity detection}

This section presents the functions of package \textbf{multiColl} \cite{multiCollR}, which implements in \textbf{R} the calculation of the correlation matrix of a model's independent variables and its determinant, the variance inflation factors, the condition number (with and without the intercept) and the Stewart index, among others.
The package also includes functions to quantify the variation of the estimates of the model's coefficients when observations are modified.

The application of the package is illustrated by an example of Henri Theil's textile consumption data shown in Table \ref{theil}. The data of Theil \cite{Theil:1971} are a time series for the period from 1923 to 1939 (17 observations) for the consumption of textiles in the Netherlands. The variables are year, volume of textile consumption per capita (base 1925=100), real income per capita (base 1925=100) and relative price of textiles (base 1925=100). We have included a dummy variable to distinguish between the 20s and the 30s.

\begin{table}
    \centering
    \caption{Henri Theil's textile consumption data} \label{theil}
    \begin{tabular}{ccccc}
        \hline\noalign{\smallskip} 
        Year & Consumption & Income & Relprice & Twenties \\
        \noalign{\smallskip}\hline\noalign{\smallskip} 
        1923 & 99.2 & 96.7 & 101.0 & 1 \\
        1924 & 99.0 & 98.1 & 100.1 & 1 \\
        1925 & 100.0 & 100.0 & 100.0 & 1 \\
        1926 & 111.6 & 104.9 & 90.6 & 1 \\
        1927 & 122.2 & 104.9 & 86.5 & 1 \\
        1928 & 117.6 & 109.5 & 89.7 & 1 \\
        1929 & 121.1 & 110.8 & 90.6 & 1 \\
        1930 & 136.0 & 112.3 & 82.8 & 0 \\
        1931 & 154.2 & 109.3 & 70.1 & 0 \\
        1932 & 153.6 & 105.3 & 65.4 & 0 \\
        1933 & 158.5 & 101.7 & 61.3 & 0 \\
        1934 & 140.6 & 95.4 & 62.5 & 0 \\
        1935 & 136.2 & 96.4 & 63.6 & 0 \\
        1936 & 168.0 & 97.6 & 52.6 & 0 \\
        1937 & 154.3 & 102.4 & 59.7 & 0 \\
        1938 & 149.0 & 101.6 & 59.5 & 0 \\
        1939 & 165.5 & 103.8 & 61.3 & 0 \\
        \noalign{\smallskip}\hline 
    \end{tabular}
\end{table}

The application of the package will also be illustrated with the dataset used by Klein and Goldberger \cite{KleinGoldberger:1964} shown in Table \ref{KG}. This dataset facilitates the analysis of consumption and wage incomes in the United States from 1936 to 1952 (data from 1942 to 1944 are unavailable due to the war). The variables are consumption, wage incomes, non-farm incomes and farm incomes.

\begin{table}
    \begin{center}
        \caption{Klein and Goldberger's dataset regarding consumption and incomes in the United States} \label{KG}
        \begin{tabular}{ccccc}
            \hline\noalign{\smallskip} 
            Year & Consumption & Wage Income & Non-farm income & Farm income \\
            \noalign{\smallskip}\hline\noalign{\smallskip} 
            1936 & 62.8 & 43.41 & 17.1 & 3.96 \\
            1937 & 65 & 46.44 & 18.65 & 5.48 \\
            1938 & 63.9 & 44.35 & 17.09 & 4.37 \\
            1939 & 67.5 & 47.82 & 19.28 & 4.51 \\
            1940 & 71.3 & 51.02 & 23.24 & 4.88 \\
            1941 & 76.6 & 58.71 & 28.11 & 6.37 \\
            1945 & 86.3 & 87.69 & 30.29 & 8.96 \\
            1946 & 95.7 & 76.73 & 28.26 & 9.76 \\
            1947 & 98.3 & 75.91 & 27.91 & 9.31 \\
            1948 & 100.3 & 77.62 & 32.3 & 9.85 \\
            1949 & 103.2 & 78.01 & 31.39 & 7.21 \\
            1950 & 108.9 & 83.57 & 35.61 & 7.39 \\
            1951 & 108.5 & 90.59 & 37.58 & 7.98 \\
            1952 & 111.4 & 95.47 & 35.17 & 7.42 \\
            \noalign{\smallskip}\hline 
        \end{tabular}
    \end{center}
\end{table}

\subsection{Correlation matrix and its determinant: function \textit{RdetR}}

The correlation coefficient measures the linear relation between two quantitative variables, and for this reason, it should not be considered if one of the variables is nonquantitative (for example, a dummy variable) or is constant. Thus, this measure is appropriate for detection of near essential multicollinearity if there is a relation between two variables. In other words, if multicollinearity is caused by a relation between more than two variables, the correlation coefficient will be unable to detect multicollinearity. As shown by García et al. \cite{Garcia:2018}, the existence of a simple correlation coefficient higher than $\sqrt{0.9} = 0.9486833$ implies the existence of problematic near essential multicollinearity between two variables. 

Regarding the determinant of the correlation matrix, a value close to zero indicates that the coefficients of simple correlation are close to 1. García et al. \cite{Garcia:2018} show that values of the determinant of the correlation matrix lower than $0.1013 + 0.00008626 \cdot n - 0.01384 \cdot k$ indicate the presence of problematic near essential multicollinearity. 
This measure may detect relations broader than that between two variables since it considers various correlation coefficients.

The following results are obtained for the considered examples:

\begin{example}
> RdetR(theil.X, T, pos = 4)
$`Correlation matrix`
         income    relprice
income   1.0000000 0.1788467
relprice 0.1788467 1.0000000

$`Correlation matrix's determinant`
[1] 0.9680139

> RdetR(KG.X)
$`Correlation matrix`
                wage.income non.farm.income farm.income
wage.income     1.0000000   0.9431118       0.8106989
non.farm.income 0.9431118   1.0000000       0.7371272
farm.income     0.8106989   0.7371272       1.0000000

$`Correlation matrix's determinant`
[1] 0.03713592
\end{example}

It is worth noting that the model must have at least two quantitative variables, and if the model has nonquantitative variables, their positions should be indicated in the design matrix\footnote{ $theil.X$ and $KG.X$ refer to the design matrixes corresponding to models in Tables \ref{theil} and \ref{KG}, respectively.} $\mathbf{X}$ for elimination in the analysis. The first column of $\mathbf{X}$ corresponding to the intercept is also excluded.

In the first case, the presence of problematic near multicollinearity is not detected with any measure. However, in the second example, although none of the correlation coefficients is higher than 0.9486833, the determinant is lower than $0.1013+0.00008626 \cdot 14-0.01384 \cdot 3 = 0.06098764$. Thus, in this case, a problematic essential near multicollinearity has been detected.

\subsection{Variance Inflation Factors: function \textit{VIF}}

García et al. \cite{Garcia:2015}, among others, shows that the main diagonal of the inverse of the correlation matrix is composed of the so-called variance inflation factors (VIFs).
Thus, the VIFs are not appropriate for nonquantitative variables. Consequently, if a model includes a nonquantitative variable, the model is required to indicate such a variable's position in the design matrix $\mathbf{X}$. The first column of $\mathbf{X}$ corresponding to the intercept is also eliminated, and the model needs to include at least two quantitative variables.

The following results are obtained for the considered examples:

\begin{example}
> VIF(theil.X, T, pos = 4)
 income relprice
1.033043 1.033043
> VIF(KG.X)
 wage.income non.farm.income farm.income
 12.296544 9.230073 2.976638
\end{example}

To interpret these results, it is convenient to take into account that the VIFs ignore the intercept, as shown, for example, by Salmerón et al. \cite{Salmeron:2018} or Galmacci \cite{Galmacci1996}. The latter also stated: ``We know that VIFs are the main diagonal elements of the correlation matrix regressors; therefore, they do not incorporate information about the constant term of the model. If collinearity involves the intercept, we cannot expect correct information from the VIF annalysis''. Thus, based on the VIFs, it is not possible to detect near nonessential multicollinearity but is only possible to detect essential multicollinearity.

Considering that values of VIFs higher than 10 indicate problematic near multicollinearity, it is possible to conclude that in the first example there is no multicollinearity, while the second example presents problematic multicollinearity.

\subsection{Condition Number: functions \textit{CN} and \textit{CNs}}

The condition number (CN) is obtained from the square root of the ratio between the maximum and the minimum eigenvalues of matrix $\mathbf{X}^{t}\mathbf{X}$ after the transformation of columns of matrix $\mathbf{X}$ to be of unit length. Values between 20 and 30 indicate a moderate near multicollinearity, while values higher than 30 indicate problematic near multicollinearity (see Belsley \cite{Belsley:1991}). Belsley \cite{Belsley1991} provides a detailed guidance of this collinearity diagnostic.

The CN considers the intercept and the possible existence of dummy variables; consequently, this measure is appropriate for detection of essential and nonessential multicollinearity.
In both examples, the results indicate that there is a problematic degree of multicollinearity.
\begin{example}
> CN(theil.X)
[1] 53.39671
> CN(KG.X)
[1] 35.88644
\end{example}

Recall that in the first example, VIFs did not detect essential near collinearity; hence, it is possible to conclude that the near multicollinearity detected by the CN should be nonessential. However, in the second example it is impossible to distinguish as clearly the kind of collinearity existing in the model.

Thus, it could be interesting to calculate the CN with and without consideration of the intercept and to calculate the increase caused by a change from not accounting for the intercept to accounting for it.

The following results are obtained for the considered examples:

\begin{example}
> CNs(theil.X)
$`Condition Number without intercept`
[1] 24.15423

$`Condition Number with intercept`
[1] 53.39671

$`Increase (in percentage)`
[1] 54.76458

> CNs(KG.X)
$`Condition Number without intercept`
[1] 30.2987

$`Condition Number with intercept`
[1] 35.88644

$`Increase (in percentage)`
[1] 15.57062
\end{example}

In the first case, the CN without the intercept indicates a moderate near multicollinearity, and an increase of 54.76458\% is observed when accounting for the intercept. In the second example, the CN without the intercept indicates a problematic near collinearity. In addition, the increase when considering the intercept is 15.57062\%, and consequently, it appears that the intercept does not have a relevant role in the existing problematic multicollinearity\footnote{Unfortunately, there is no threshold for determining whether the percentage increase will imply a problematic near nonessential multicollinearity.}.

\subsection{Stewart index: function \textit{ki}}

Stewart \cite{Stewart:1987} presented Stewart's index, $k_{i}^{2}$, which measures the relation between the i-th column of $\mathbf{X}$ and the rest of columns of $\mathbf{X}$. Although Stewart identified this index with the VIF, these measures only coincide if the variables to which the measures are applied have zero mean.

Thus, for $i \geq 2$, $k_{i}^{2} = VIF(i) + a_{i}$, where $VIF(i)$ is the VIF associated with the i-th variable of $\mathbf{X}$, and $a_{i}$ is a value that depends on the average of the i-th variable of $\mathbf{X}$, which equals zero if the mean is zero.

Because this expression depends on the VIF, it is not appropriate to calculate this measure for dummy variables. In addition, as the VIF is only able to detect essential near multicollinearity, the ratio $VIF(i)/k_{i}^{2}$ can be interpreted as the percentage of essential near multicollinearity existing in the i-th variable of $\mathbf{X}$. The difference from 100\% will be the percentage of nonessential near multicollinearity.

Another aspect that distinguishes this measure from the VIF is that it can be calculated for $i=1$, that is, for the intercept. In this case, it measures the nonessential multicollinearity existing in the model.

Finally, note that there is no threshold for Stewart's index that can be used to establish that the model is characterized by problematic near multicollinearity.

The following results are obtained for the considered examples:

\begin{example}
> ki(theil.X, T, pos = 4)
$`Stewart index`
[1] 403.20963 415.28266 23.50258

$`Proportion of essential collinearity in i-th independent variable
    (without intercept)`
[1] 0.2487566 4.3954455

$`Proportion of non-essential collinearity in i-th independent variable
    (without intercept)`
[1] 99.75124 95.60455

> ki(KG.X)
$`Stewart index`
[1] 17.86327 185.96422 156.50013 39.16836

$`Proportion of essential collinearity in i-th independent variable
    (without intercept)`
[1] 6.612317 5.897805 7.599598

$`Proportion of non-essential collinearity in i-th independent variable
    (without intercept)`
[1] 93.38768 94.10219 92.40040
\end{example}

In the first example, problematic near (non-essential) multicollinearity is detected. In addition, the second explanatory variable seems to be  closely related to the intercept due to the 99.75124\% of the near multicollinearity caused by this variable is non-essential.

The second example presents also a high percentage of non-essential multicollinearity but also the essential collinerity is problematic regardless of the percentage. In addition, the value of the Stewart's index associated to the intercept is not very high (compared to the first example). Anyway, the existence of a threshold for this measure will help to obtain conclusions.

\subsection{A special case, the simple linear regression model: function \textit{SLM}}

The simple linear model (model (\ref{model0}) with $k=2$) is a particular case. The relevance of this case is due to it is systematically ignored in many different statistical softwares to determine if the near multicollinearity is problematic. For the best of our knowledge, the existing packages in \textbf{R} do not calculate any measure to detect near multicollinearity in this case as in the simple linear model there is no collinearity.

However, this paper shows that a simple linear model can present non-essential collinearity. In this case, it is not appropriate to calculate the VIF\footnote{Salmerón et al. \cite{Salmeron:2018} shows that in the simple linear regression model the VIF is always equal to 1.} and is not possible to obtain the matrix of simple correlations. Only the CN and the Stewart's index seem appropriate measures.

In addition, the independent variable should present very little variability to be related to the intercept. For this reason, the calculation of its coefficient of variation can be very useful as is indicated by Simon and Lesage \cite{SimonLesage1988}: ``the coefficient of variation may be useful in combination with one of the other diagnostics which is only sensitive to the second (non-essential) type of ill-conditioning''.
The application of the coefficient of variation was also proposed by Gunst \cite{Gunst1984}: ``a coefficient of variation small calls for immediate investigation of collinearity''. Recently, Salmerón et al. \cite{Salmeron2019b} proposed to conclude that there is non-essential worrying collinearity if the coefficient of variation of an independent variable is lower than  0.1002506. It is to say, when the independent variable is highly correlated to the intercept.

In the case of a dummy independent variable, the proportion of ones could be a good approximation to know the relation of this variable with the intercept.

Considering the following simple linear regression for data in Table \ref{theil}:
$$consumption = f(income), \quad consumption = f(relprice), \quad consumption = f(twentys),$$
the following results are obtained:
\begin{example}
> SLM(theil.X[,1:2])
$`Coeficient of Variation`
[1] 0.04993766

$`Variance Inflation Factor`
[1] 1

$`Condition Number`
[1] 40.07489

$`Stewart index`
[1] 401.9994 401.9994

> SLM(theil.X[,c(1,3)])
$`Coeficient of Variation`
[1] 0.2144185

$`Variance Inflation Factor`
[1] 1

$`Condition Number`
[1] 9.43356

$`Stewart index`
[1] 22.75082 22.75082

> SLM(theil.X[,c(1,4)], T)
$`Proportion of ones in the dummy variable`
[1] 41.17647

$`Condition Number`
[1] 2.140501

> SLM(KG.X)
[1] "Only 2 independent variables are needed (including the intercept)"
\end{example}

In the first model, the value of CN indicates that the near (nonessential) multicollinearity is problematic. This value is accompanied by a value of CV that is very close to zero and very high values of Stewart's index. In the second model, the values of CV and Stewart's index are moderate. In addition, the value of CN is less than the established threshold.

Finally, the third model is also characterized by a value of CN less than the established threshold, and the proportion of ones existing in the dummy variable also indicates a slight linear relation with the intercept.

\subsection{Summary of detection measures: function \textit{multiCol}}

All the previous functions can be collected in the following unified function whose results coincide with the results previously shown:

\begin{example}
> multiCol(theil.X, T, pos = 4)
$`Coeficients of Variation`
[1] 0.04993766 0.21441845

$`Proportion of ones in the dummys variable`
[1] 41.17647

$`R and det(R)`
$`R and det(R)`$`Correlation matrix`
         income    relprice
income   1.0000000 0.1788467
relprice 0.1788467 1.0000000

$`R and det(R)`$`Correlation matrix's determinant`
[1] 0.9680139

$`Variance Inflation Factors`
 income relprice
1.033043 1.033043

$CN
$CN$`Condition Number without intercept`
[1] 24.15423

$CN$`Condition Number with intercept`
[1] 53.39671

$CN$`Increase (in percentage)`
[1] 54.76458

$ki
$ki$`Stewart index`
[1] 403.20963 415.28266 23.50258

$ki$`Proportion of essential collinearity in i-th independent variable
    (without intercept)`
[1] 0.2487566 4.3954455

$ki$`Proportion of non-essential collinearity in i-th independent variable
    (without intercept)`
[1] 99.75124 95.60455


> multiCol(theil.X[,1:2])
$`Coeficient of Variation`
[1] 0.04993766

$`Variance Inflation Factor`
[1] 1

$`Condition Number`
[1] 40.07489

$`Stewart index`
[1] 401.9994 401.9994


> MultiCol(theil.X[,c(1,3)])
$`Coeficient of Variation`
[1] 0.2144185

$`Variance Inflation Factor`
[1] 1

$`Condition Number`
[1] 9.43356

$`Stewart index`
[1] 22.75082 22.75082


> MultiCol(theil.X[,c(1,4)], T)
$`Proportion of ones in the dummy variable`
[1] 41.17647

$`Condition Number`
[1] 2.140501

> MultiCol(KG.X)
$`Coeficients of Variation`
[1] 0.2660921 0.2503487 0.2867863

$`Proportion of ones in the dummys variable`
[1] "At least one qualitative independent variable are needed
    (excluding the intercept)"

$`R and det(R)`
$`R and det(R)`$`Correlation matrix`
                wage.income non.farm.income farm.income
wage.income     1.0000000   0.9431118       0.8106989
non.farm.income 0.9431118   1.0000000       0.7371272
farm.income     0.8106989   0.7371272       1.0000000

$`R and det(R)`$`Correlation matrix's determinant`
[1] 0.03713592

$`Variance Inflation Factors`
 wage.income non.farm.income farm.income
 12.296544 9.230073 2.976638

$CN
$CN$`Condition Number without intercept`
[1] 30.2987

$CN$`Condition Number with intercept`
[1] 35.88644

$CN$`Increase (in percentage)`
[1] 15.57062

$ki
$ki$`Stewart index`
[1] 17.86327 185.96422 156.50013 39.16836

$ki$`Proportion of essential collinearity in i-th independent variable
    (without intercept)`
[1] 6.612317 5.897805 7.599598

$ki$`Proportion of non-essential collinearity in i-th independent variable
    (without intercept)`
[1] 93.38768 94.10219 92.40040
\end{example}

\section{Perturbation: functions \textit{perturb} and \textit{perturb.n}}

Two types of consequences of problematic near multicollinearity can be distinguished: the numerical instability in estimating the coefficients of independent variables and the obtention of inflated variances of the estimated coefficients with well-known implications for the individual inference of the model.

The following results are obtained by estimating both models by OLS:

\begin{example}
> reg.theil = lm(consume~income+relprice+twentys)
> summary(reg.theil)

Call:
lm(formula = consume ~ income + relprice + twentys)

Residuals:
 Min 1Q Median 3Q Max
-9.372 -4.122 1.021 3.504 9.408

Coefficients:
           Estimate Std. Error t value Pr(>|t|)
Intercept) 126.1695 28.4634    4.433   0.000676 ***
income     1.0308   0.2760     3.735   0.002497 **
relprice   -1.2574  0.2062     -6.097  3.8e-05 ***
twentys    -4.5355  6.7753     -0.669  0.514947
---
Signif. codes: 0 ‘***’ 0.001 ‘**’ 0.01 ‘*’ 0.05 ‘.’ 0.1 ‘ ’ 1

Residual standard error: 5.676 on 13 degrees of freedom
Multiple R-squared: 0.9529, Adjusted R-squared: 0.942
F-statistic: 87.68 on 3 and 13 DF, p-value: 7.048e-09

> reg.KG = lm(consumption ~ wage.income + non.farm.income
    + farm.income)
> summary(reg.KG)

Call:
lm(formula = consumption ~ wage.income + non.farm.income
    + farm.income)

Residuals:
 Min 1Q Median 3Q Max
-13.494 -1.847 1.116 2.541 6.460

Coefficients:
                Estimate Std. Error t value Pr(>|t|)
(Intercept)     18.7021  6.8454     2.732   0.0211 *
wage.income     0.3803   0.3121     1.218   0.2511
non.farm.income 1.4186   0.7204     1.969   0.0772 .
farm.income     0.5331   1.3998     0.381   0.7113
---
Signif. codes: 0 ‘***’ 0.001 ‘**’ 0.01 ‘*’ 0.05 ‘.’ 0.1 ‘ ’ 1

Residual standard error: 6.06 on 10 degrees of freedom
Multiple R-squared: 0.9187, Adjusted R-squared: 0.8943
F-statistic: 37.68 on 3 and 10 DF, p-value: 9.271e-06
\end{example}

In the first case, there is no apparent contradiction. However, in the second case, the null hypothesis of the individual significance test is not rejected (at 5\% significance level), while the global significance test indicates that at least one coefficient is different from zero. This contradiction is a symptom of problematic near multicollinearity and serves as evidence of the negative effects of collinearity on the statistical analysis of the model.

An interesting question is how to measure the consequences of instability of the coefficients' estimates. Belsley \cite{Belsley:1982} showed how to quantify whether small changes in the data imply relevant changes in the estimates. In particular, if we consider the following version of vector $\mathbf{x}$ perturbed by 1\%,
$$\mathbf{x}_{p} = \mathbf{x} + 0.01 \cdot \mathbf{r} \cdot \frac{||\mathbf{x}||}{||\mathbf{r}||},$$
where $\mathbf{r}$ is a random vector with dimension equal to that of $\mathbf{x}$,
and $||\mathbf{x}|| = \sqrt{\sum \limits_{i=1}^{n} x_{i}^{2}}$, the variation experienced by the coefficients' estimates will be quantified by the following expression:

$$\frac{||\boldsymbol{\beta} - \boldsymbol{\beta}_{p}||}{||\boldsymbol{\beta}||},$$
where subindex $p$ denotes the estimates of the perturbed model.

Considering that only quantitative variables can be perturbed in this way, it is necessary to specify the observations of the model, $(\mathbf{y}, \mathbf{X})$, and the positions (without the intercept) where the quantitative variables are placed in the design matrix $\mathbf{X}$. Additionally, it is necessary to indicate how many times the perturbation of the model is to be repeated.

With 5000 iterations, the following results are obtained:

\begin{example}
> tol = 0.01
> media = 10
> dv = 10
> iteraciones = 5000

> perturb.n.T = perturb.n(theil.y.X, iteraciones, media, dv, tol, pos = c(1,2))
> c(mean(perturb.n.T[,1]), sd(perturb.n.T[,1]))
[1] 1.000000e+00 7.496194e-16
> c(mean(perturb.n.T[,2]), sd(perturb.n.T[,2]))
[1] 4.132412 2.815622
> c(min(perturb.n.T[,2]), max(perturb.n.T[,2]))
[1] 0.02346593 17.83920908
> c(quantile(perturb.n.T[,2], prob=0.025), quantile(perturb.n.T[,2], prob=0.975))
 2.5% 97.5%
 0.5230459 10.9377709

> perturb.n.KG = perturb.n(KG.y.X, iteraciones, media, dv, tol, pos = c(1,2,3))
> c(mean(perturb.n.KG[,1]), sd(perturb.n.KG[,1]))
[1] 1.000000e+00 7.648695e-16
> c(mean(perturb.n.KG[,2]), sd(perturb.n.KG[,2]))
[1] 3.048448 1.798213
> c(min(perturb.n.KG[,2]), max(perturb.n.KG[,2]))
[1] 0.05071945 10.68055642
> c(quantile(perturb.n.KG[,2], prob=0.025), quantile(perturb.n.KG[,2], prob=0.975))
 2.5% 97.5%
0.544167 6.989646
\end{example}

The function returns two vectors: the first vector refers to the perturbation introduced in the quantitative independent variables and the second refers to the percentage of change experienced by the estimates of the model's coefficients.

In both cases, note that the introduced perturbation is indeed 1\%. Different measures are calculated to obtain the percentage changes in the estimates of the model's coefficients. In no cases do small changes in the data lead to significant changes in the coefficients' estimates. In other words, although collinearity is problematic, its consequences do not affect the numerical stability of the model. Once again, it could be convenient to have a threshold for the percentage change, based on which it would be determined whether the multicollinearity is problematic or not.

Note that in the first model, there is no negative effect either on the numerical analysis or on the statistical analysis; consequently, a more appropriate model would not make a difference. This statement is based on the idea that the presence of multicollinearity favors the prediction of the model if the new information presents the same linear structure of the observations of the independent variables as that used to estimate the model (see, for example, Gujarati \cite{Gujarati:2003}).

Finally, note that package \textbf{perturb} \cite{perturb2012} allows the evaluation of how small changes in observations affect to the estimations providing the mean, standard deviation, minimum and maximum of the parameter estimates. This package also allows the perturbation of categorial variable. Under our humble consideration, this possibility does not seem to make sense.

\section{Conclusions}

This paper presents some of the functions of the package \textbf{multiColl} to obtain the correlation matrix of the model's independent variables and its determinant (function \textit{RdetR}), the variance inflation factors (function \textit{VIF}), the condition number with and without intercept (functions \textit{CN} and \textit{CNs}), the Stewart index (function \textit{ki}), the coefficient of variation  (function \textit{CV}) and the variations in the estimations of coefficients after small changes in the data  (functions \textit{perturb} and \textit{perturb.n}). This package also allows the diagnosis of nonessential multicollinearity in a simple linear model (function \textit{SLM}).


%
%

\bibliographystyle{spmpsci}      
\bibliography{References}   

\end{document}